\documentclass[letter]{aa} 
\usepackage{graphicx}
\usepackage{txfonts}

\begin{document}
   \title{Transient jets in V617 Sagittarii
	\thanks{Data obtained at Observat\'orio do Pico dos Dias, LNA/MCT,
	        Brazil.}
	}

   \author{J. E. Steiner \inst{1}, A. S. Oliveira \inst{2,3}, 
  	C. A. O. Torres \inst{4} \and
  	A. Damineli \inst{1}
          }

   \institute{Instituto de Astronomia, Geof\'{\i}sica e
	 Ci\^encias Atmosf\'ericas, Universidade de S\~ao Paulo,
	  05508-900, S\~ao Paulo, SP, Brazil\\
             \email{steiner@astro.iag.usp.br}
	     \and
	IP\&D, Universidade do Vale do Para\'{\i}ba, Av. Shishima
	 Hifumi, 2911, CEP 12244-000, S\~ao Jos\'e dos Campos, SP,
	  Brazil \\
	     \email{alexandre@univap.br}
	     \and
	      SOAR Telescope, Casilla 603, La Serena, Chile
	      \and Laborat\'orio Nacional de Astrof\'{\i}sica -
	       LNA/MCT, Itajub\'a, MG, Brazil
             }

   \date{Received / Accepted}

  \abstract
   {Some of the luminous Compact Binary Supersoft X-Ray sources (CBSS) have shown indications of jets, also called satellites due to their appearance in the spectra. In V Sagittae (V~Sge) stars, the galactic counterparts of the CBSS, such features have been reported only for \object{WX~Cen}.}
   {If V~Sge stars are indeed the analogs of CBSS, one may expect transient jet emission in other objects of this class.}
   {Spectroscopic observations of the V~Sge star \object{V617~Sgr} have been made, both at high photometric state and at decline.}
   {We show that \object{V617~Sgr} presents H$\alpha$ satellites at high photometric state with velocities of $\pm 780$~km~s$^{-1}$.}
   {This feature confirms, once more, the CBSS nature of the V~Sge stars, however the details of the spectral characteristics also suggest that the two groups of stars display some intrinsic spectroscopic differences, which are likely to be due to a selection effect related to chemical abundance.}

   \authorrunning{J. E. Steiner et al.}
   \titlerunning{Transient jets in V617 Sagittarii}
   \maketitle
%

\section{Introduction}

   The CBSS are a group of luminous stars with high temperatures ($20 - 80$~eV) discovered in the Magellanic Clouds (Tr\"umper et al. \cite{trump}). 
Only two galactic sources, \object{QR~And} and \object{MR~Vel}, have been classified as CBSS on the grounds of  their supersoft X-ray emission. However, the V~Sge stars as a group, although not supersoft X-ray emitters, have been proposed to be the galactic counterpart of the CBSS (Steiner \& Diaz \cite{stei98}).
Why then, are V~Sge stars not detected as luminous supersoft sources? In part, because the Supersoft X-ray photons might be absorbed in the intergalactic medium but, perhaps also because the chemical environment is distinct, both in the interstellar medium and in the stars themselves. The similarities between the two groups of objects are related to their spectroscopic appearance (\ion{O}{VI}, \ion{N}{V} and H Balmer emission lines; large \ion{He}{II}/H$\beta$ ratio) as well as to the similarities observed in their light curves. Optical absolute magnitudes  are comparable as are their orbital periods.  Some stars from both groups present high/low optical photometric states. 

The origin of the luminosity of these objects seems to be related to hydrostatic nuclear burning on the surface of the white dwarf (van den Heuvel et al. \cite{heuvel}). This may happen in a binary system when the accretion rate onto the white dwarf is high -- about $10^{-7}$ M$_{\sun}$~yr$^{-1}$. Two causes have been proposed for such an event to occur. When the mass ratio is inverted as compared to typical cataclysmic variables ($M_2 > M_1$), a dynamical instability takes place and this induces a mass transfer on the Kelvin-Helmholtz time-scale (van den Heuvel et al. \cite{heuvel}). 
Alternatively, the low-mass secondary may be
strongly irradiated by the supersoft source and as a consequence, expands, forcing a high mass transfer (van Teeseling \& King \cite{tees}). These are the Wind Driven Mass Transfer -- WDMT -- systems. 

Are there differences between the CBSS and V~Sge stars, other than the detection of Supersoft X-rays? If not, then one may predict that most of the phenomena seen in one group should be detected in the other group too. For example, optical emission line satellites should be detected in the spectra of the V~Sge stars, as they are in CBSS.

Spectral emission line satellites have been detected in some but not all CBSS (Southwell et al. \cite{south}; Crampton et al. \cite{cramp}; Tomov et al. \cite{tomov}, Becker et al. \cite{becker}; Quaintrell \& Fender \cite{quaintr}; Motch \cite{motch}; Cowley et al. \cite{cow98}) indicating the presence of symmetric jets. These features are transient with timescales of months. Up to now, among the V~Sge stars, only \object{WX~Cen} has presented indications of such events (Oliveira \& Steiner \cite{oliv}).

In the present paper we report the detection of H$\alpha$ features that we identify as satellites in the star \object{V617~Sgr}. This is a V~Sge object (Steiner \& Diaz \cite{stei98}) that has an orbital period of 4.97 hours (Steiner et al. \cite{stei99}). It has recently been shown to have an increasing orbital period and is therefore, a wind driven accreting CBSS (Steiner et al. \cite{stei06}).

\section{Observations and data analysis}

A spectrum of V617~Sgr at high state, covering the range from $6200~{\AA}$ to $6800~{\AA}$ with a spectral resolution of $1.7~{\AA}$ FWHM, was obtained with the Eucalyptus spectrograph on May 21, 2003 at the 1.6 m telescope in the Observat\'orio do Pico dos Dias -- OPD, operated by MCT/LNA in Itajub\'a, Brazil. 
The Eucalyptus (de Oliveira et al. \cite{eucalyp}) is a prototype of the SIFS -- SOAR Integral Field Spectrograph. It is composed of an array of 16x32 50 $\mu$m fibers that covers a field
 of 15x30 arcsec$^2$ on sky with an scale of 0.93 arcsec per pixel. The detector used was a Marconi 
 2048x4608 pixels back-illuminated CCD with 13.5 $\mu$m$^2$ per pixel. 
On July 1 -- 3, 2003, six additional spectra of V617~Sgr were taken with the Coud\'e spectrograph at the same telescope, three of them covering from $3820~{\AA}$ to $4930~{\AA}$ and the remaining spectra covering from $5830~{\AA}$ to  $6950~{\AA}$. 
The Coud\'e spectrograph was employed with a 600 l mm$^{-1}$ grating and the same Marconi 2048x4604 CCD as on May 21, 2003, resulting in a $0.6~{\AA}$ FWHM spectral resolution. All spectra were corrected for bias and flatfield signatures, and wavelength calibration was performed with the aid of spectra of calibration lamps. The data reduction was executed with the standard procedures, using IRAF~\footnote{IRAF is distributed by the National Optical Astronomy Observatories,
which are operated by the Association of Universities for Research in Astronomy, Inc., under cooperative agreement
with the National Science Foundation.} routines.

Photometric monitoring of the source was performed simultaneously with all spectroscopic observations described above. These photometric observations were done with the Zeiss 60 cm telescope of OPD and the thin back-illuminated EEV CCD 002-06 and a Wright Instruments thermoelectrically cooled camera. The images were obtained through the Johnson V filter and corrected for bias and flatfield. Differential aperture photometry was carried out with the DAOPHOT II package.

\begin{table*}[!ht]
\caption{Spectral properties for CBSS and V~Sge stars with known orbital periods.}
\label{tab}
\begin{flushleft}
\begin{tabular}{lllllllllll}
\hline
\hline\noalign{\smallskip\smallskip}
            \noalign{\smallskip}
  Source   & Type  & P$_{orb}$ &  High/low & Emission         & Absorption        & $-W{_\lambda}$(\ion{He}{II}) &i & FWHM      & $\Delta$v jet & Refs.$^{\mathrm{a}}$\\
           &       &   (days)  &  states   & jets (km s$^{-1}$)& jets (km s$^{-1}$)&  ({\AA})                     &($\degr$)  & (km~s$^{-1}$)&  (km s$^{-1}$) & \\
\hline
\noalign{\smallskip}
RX J0537.7-7034	& CBSS	&0.145	&...	&...	&...	&2.9--4.8	&45--70	&...	&...	&1\\
1E0035.4-7230& CBSS	&0.171	&...	&...	&...	&2.6	&20--50	&700	&...	&2\\
V617~Sgr& V~Sge	&0.207	&y	&780	&...	&35--113	&72	&710--2900	&$\sim600$	&3, 4\\
WX~Cen	& V~Sge	&0.416	&...	&3500	&-2900	&12--37	&55	&700--1100	&...	&5, 6\\
CAL~87	& CBSS	&0.442	&...	&...	&...	&14	&77	&790	&...	&7, 8\\
V~Sge	& V~Sge	&0.514	&y	&...	&...	&46--117&71	&700	&...	&9, 10\\
QR~And	& CBSS	&0.660	&y	&815	&-700	&14	&56	&470--650	&400	&11, 12\\
RX~J0513.9-6951	& CBSS	&0.762	&y	&3900	&-3458	&22	&15	&370	&...	&13\\
CAL 83	& CBSS	&1.041	&y	&...	&...	&12	&20--30	&350	&...	&13\\
MR~Vel	& CBSS	&4.028	&...	&5200	&...	&5.3	&30--40	&330--650	&...	&14\\
\noalign{\smallskip}
\hline
\end{tabular}
\end{flushleft}
\smallskip\noindent
\begin{list}{}{}
\item[$^{\mathrm{a}}$] References: 1- Greiner et al. \cite{greiner}; 2- van Teeseling et al. \cite{teesetal}; 3- Steiner et al. \cite{stei99}; 4- Steiner et al. \cite{stei06}; 5- Diaz \& Steiner \cite{diaz95}; 6- Oliveira \& Steiner \cite{oliv}; 7- Hutchings et al. \cite{hutchings}; 8- Schandl et al. \cite{schandl}  9- Herbig et al. \cite{herbig}; 10- Smak et al. \cite{smak}; 11- McGrath et al. \cite{mcgrath}; 12- Deufel et al. \cite{deufel}; 13-  Cowley et al. \cite{cow98}; 14- Motch \cite{motch}.
\end{list}
\end{table*}

\begin{figure}
      \includegraphics[width=0.5\textwidth,clip]{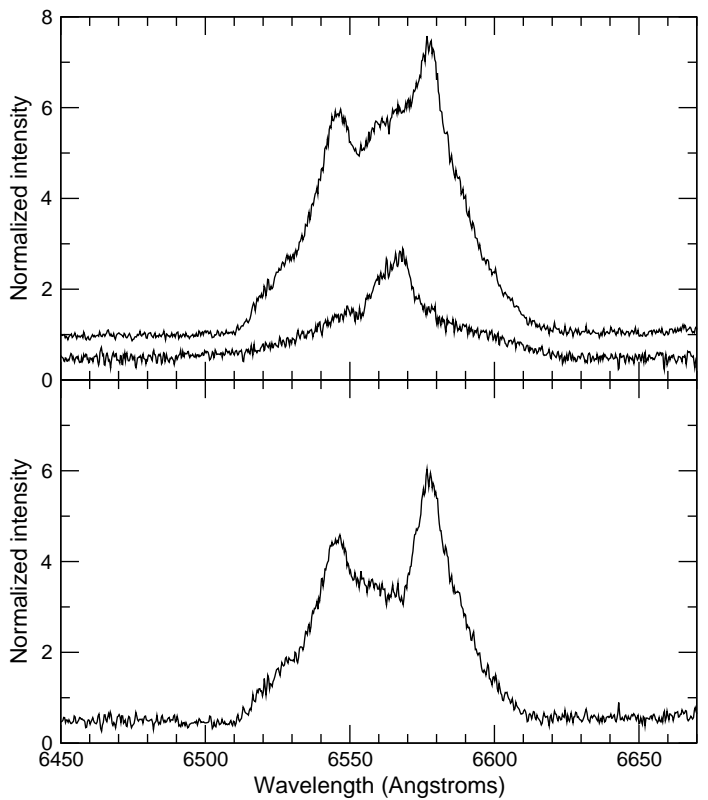}
     \caption{\textit{Top panel:} The H$\alpha$ profile of V617~Sgr at high state (upper spectrum), taken with 
     the Eucalyptus spectrograph, and the same spectral line at decline (lower spectrum), taken with the Coud\'e
     spectrograph six weeks later. \textit{Bottom panel:} The difference between the high state spectrum and the
      decline spectrum at H$\alpha$.}
     \label{spec}
  \end{figure}

V617~Sgr has a median $V$ magnitude of $V=14.7$ but on occasions it brightens by at least $\Delta m=1.6$ (Steiner et al. \cite{stei99}). On May 21, 2003 the star was at photometric high state, as shown by the simultaneous spectroscopic observations and CCD photometry, since its $V$ magnitude was about $V\sim13.1$; on that occasion, H$\alpha$ clearly showed an anomalous profile, displaying a ``double horn'' -- see Fig.~\ref{spec}. About one month later, on July
 1 -- 3, the star had faded to about $V\sim13.9$, i.e. 0.8 magnitudes fainter, and the two
  horns disappeared completely. 

We subtracted the two spectra, at maximum and at decline, and obtained a profile in which the two horns are quite conspicuous (Fig.~\ref{spec}). These horns are somewhat similar to the detached satellites seen in CBSS, where are interpreted as originating from jets.  Before the subtraction each spectrum was normalized relative to the flux ratio $F_{high}/F_{decline}=2.1$ derived from the simultaneous photometry. 
In the difference spectrum one can clearly see the two satellites. The blue component has a velocity of $v = -790$ km s$^{-1}$ and the red component, $v = +770$ km s$^{-1}$.

We measured the width of the satellites as FWHM $\sim600$ km s$^{-1}$; this measurement has significant uncertainty due to the blending with the H$\alpha$ main component, which is very broad at high state.
At high state, the FWHM of H$\alpha$ was enhanced when compared to the low state; our measurements indicate that, for H$\alpha$, FWHM$=1650$ km s$^{-1}$ at high state, while at the decline  FWHM$=710$ km s$^{-1}$. Steiner et al. (\cite{stei99}) reported observations of H$\alpha$ at high state, shown in their Fig. 5. Its profile showed two peaks, less prominent when compared to those reported in this paper. On that occasion, H$\alpha$ had FWHM$= 2900$ km s$^{-1}$ and the two small peaks were separated by $\pm 1050$ km s$^{-1}$.
 
The H$\alpha$ equivalent width at high state was also enhanced. We measured it as $W_{\lambda}$(H$\alpha)= 308~(\pm 8)$~{\AA} at high state and $W_{\lambda}$(H$\alpha)= 212~(\pm 15)$~{\AA} at decline, comparable to that measured by Steiner et al. (\cite{stei99}) who found  288~{\AA}; this is about 4 times larger than the value at minimum state (70~{\AA}). In other words, the flux in the H$\alpha$ line at high state was about 19 times stronger than that at quiescence. 
Our measurements of \ion{He}{II} 4686~{\AA} and H$\beta$ at decline yield $W_{\lambda}$(\ion{He}{II})= 113$~(\pm 15)$~{\AA} and $W_{\lambda}$(H$\beta)= 65~(\pm 3)$~{\AA}, which are both 3.2 times larger than the values found at low state (Cieslinski et al. \cite{cies}), i.e., $W_{\lambda}$(\ion{He}{II}) = 35~{\AA} and $W_{\lambda}$(H$\beta$)= 20~{\AA}.

\section{Discussion and conclusions} 

Do we see \textit{bona fide} jets or is it just a collimated wind? From the point of view of CBSS, there seems to be no difference between the jets in \object{V617~Sgr} and in \object{QR~And}, with the exception that the former has broader lines than any CBSS.
Some similarities with \object{QR~And} are quite striking. Emission as well as absorption events have been detected in this object (Becker et al. \cite{becker}; Tomov et al. \cite{tomov}; Quaintrell \& Fender \cite{quaintr}, Cowley et al. \cite{cow98}) with velocities that are quite similar to those reported here for \object{V617~Sgr}. The jet velocities in \object{QR~And} are $\pm815$ km s$^{-1}$ and the jet line width is 400 km s$^{-1}$. Absorption events of about -700 km s$^{-1}$ have also been reported (Cowley et al. \cite{cow98}).

The asymmetry seen in the line profile, at high state and at decline, as well as in the difference of both, is likely due to the P Cygni profile seen in other systems of this class, and in particular in QR~And (see Becker et al. \cite{becker} for a discussion).

A detailed analysis of Table~\ref{tab} suggests that there are some intrinsic differences between CBSS and V~Sge stars. First, the reported equivalent width of \ion{He}{II} is always smaller than 25~{\AA} for CBSS, while for the V~Sge stars it is usually larger than this value, being smaller occasionally only for \object{WX~Cen}. Similarly the largest line width for a CBSS is seen in \object{CAL~87}, namely, 790 km s$^{-1}$. Not by coincidence this is also the system with highest inclination. In V~Sge stars the lines are always broader than about 700 km s$^{-1}$. In conclusion, in  V~Sge stars, lines are usually stronger and broader when compared to CBSS. The only explanation we find is that this is a selection effect associated with chemical abundance. Environments and stars with low chemical abundance (CBSS) have weaker winds so supersoft X-rays can escape and not be absorbed by the interstellar
medium. This is the case of the Magellanic Clouds. The opposite happens in
stars and environmets with high chemical abundance (V Sge stars).

What is the intrinsic velocity of the jets? In order to estimate this we need to know the inclination of the orbit. A useful estimate of the inclination may be derived considering Fig. 2 from Cieslinski et al. (\cite{cies}). In their upper panel one can see that the equivalent width of \ion{He}{II} line is phase dependent, increasing near phase zero. 
This is interpreted as the emitting gas being extended in the direction perpendicular to the disc. Contrary to this, the \ion{N}{V} line does not change its equivalent width. This suggests that the line is not extended above the disc. Moreover, \ion{N}{V} being a very high ionization line, it must be emitted very close to the white dwarf. If the region of the white dwarf were to be eclipsed, this line would disappear at phase zero.   If none of the line was eclipsed, its equivalent width would increase at mid-eclipse, just like \ion{He}{II}. We conclude that the inclination angle is such that the white dwarf is nearly eclipsed. If the mass ratio is $q=0.4$ (Steiner et al. \cite{stei06}) we derive that $i=72\degr$. As $V_{obs}/V_{intr} \sim cos~i$, we obtain that the intrinsic velocity of the jets is $V_{intr} = \pm 2500$ km s$^{-1}$. This is nearly the escape velocity (Hamada \& Salpeter \cite{hama}) for a 0.5 M$_{\sun}$ white dwarf. However, Steiner et al. (\cite{stei06}) proposed a larger value, $M_1 > 1.3$ M$_{\sun}$, for the primary star, so that the intrinsic velocity must be smaller than the escape velocity of the white dwarf.

Similar to \object{QR~And}, \object{V617~Sgr} has jets with much lower velocities when compared to low inclination systems like \object{RX~J0513.9-6951} and \object{MR~Vel}. This is not surprising given the difference in inclination. \object{V617~Sgr} is the system with the shortest orbital period and highest inclination for which satellites have been detected. It is also the system with the broadest observed emission lines. These latter characteristics are relevant in the sense that they force the satellites to be blended with the main H$\alpha$ main emission, so that they look less conspicuous when compared to other systems.

Is there a relation between the presence of the jets and the high state? In other systems, no such correlation has been reported. It is, therefore, relevant to mention such a clear connection in the present case.

We conclude that the double horn seen in the V~Sge star \object{V617~Sgr} at high state is a satellite phenomenon similar to those seen in CBSS and attributed to transient jets. This is yet an additional confirmation that both  classes are basically the same phenomenon, differentiated only by their chemical abundance.

\begin{acknowledgements}
A. S. Oliveira thanks FAPESP (grant 03/12618-7) for financial support.
We would like to thank Dr. Albert Bruch for his careful reading of the manuscript.
\end{acknowledgements}

\end{document}